\documentclass[conference]{IEEEconf}
\IEEEoverridecommandlockouts

\usepackage{titlesec}
\usepackage{balance} 
\usepackage{stfloats}
\usepackage{amsthm}
\usepackage{cite}
\usepackage{amsmath,amssymb,amsfonts}
\usepackage{graphicx}
\usepackage{textcomp}
\usepackage{xcolor}
\usepackage{booktabs}
\usepackage{multirow}
\usepackage{enumitem}
\usepackage{hyperref}
\usepackage{graphicx}
\usepackage{subcaption}
\usepackage{makecell}
\usepackage[shortcuts,acronym]{glossaries}
\setkeys{glslink}{hyper=false}
\usepackage{todonotes}
\usepackage{pifont}
\usepackage{algorithm}
\usepackage[noend]{algpseudocode}
\usepackage{xspace}
\usepackage{orcidlink}

\usepackage{tikz}
    \usetikzlibrary{patterns}
\usepackage{pgfplots}
\pgfplotsset{compat=newest}

\def\BibTeX{{\rm B\kern-.05em{\sc i\kern-.025em b}\kern-.08em
    T\kern-.1667em\lower.7ex\hbox{E}\kern-.125emX}}

\theoremstyle{definition}
\newtheorem{definition}{Definition}

\newcommand{\padraig}{\textit{PADRAIG}\xspace}

\newcommand{\Atwelve}{\ensuremath{\hat{A}}\textsubscript{12}\xspace}

\newcommand{\betterNegl}{\ding{51}\xspace}
\newcommand{\betterSmall}{\ding{51}\ding{51}\xspace}
\newcommand{\betterMedium}{\ding{51}\ding{51}\ding{51}\xspace}
\newcommand{\betterLarge}{\ding{51}\ding{51}\ding{51}\ding{51}\xspace}
\newcommand{\worseNegl}{\ding{55}\xspace}
\newcommand{\worseSmall}{\ding{55}\ding{55}\xspace}
\newcommand{\worseMedium}{\ding{55}\ding{55}\ding{55}\xspace}
\newcommand{\worseLarge}{\ding{55}\ding{55}\ding{55}\ding{55}\xspace}
\newcommand{\same}{\ensuremath{\equiv}\xspace}

\newacronym{padraig}{\textit{PADRAIG}}{Precise AnDRoid Automated Input Generation}
\newacronym{aut}{AUT}{application under test}
\newacronym{api}{API}{application programming interface}
\newacronym{ui}{UI}{user interface}
\newacronym{gui}{GUI}{graphical user interface}
\newacronym{cfg}{CFG}{control flow graph}
\newacronym{ecfg}{eCFG}{extended control flow graph}
\newacronym{fsm}{FSM}{finite state machine}
\newacronym{efg}{EFG}{event flow graph}
\newacronym{atg}{ATG}{activity transition graph}
\newacronym{dfs}{DFS}{depth-first search}
\newacronym{jvm}{JVM}{Java virtual machine}
\newacronym{sdk}{SDK}{software development kit}
\newacronym{dvm}{DVM}{Dalvik virtual machine}
\newacronym{stoat}{Stoat}{STOchastic model App Tester}
\newacronym{art}{ART}{Android Runtime}
\newacronym{apk}{APK}{Android PacKage}
\newacronym{a3e}{A3E}{Automatic Android App Explorer}
\newacronym{bfs}{BFS}{breath-first search}
\newacronym{loc}{LoC}{lines of code}
\newacronym{gml}{GML}{graph modelling language}
\newacronym{smog}{STGFA-SMOG}{search-based test generation framework for Android apps with support for multi-objective generation}

\begin{document}

\title{\padraig: Precise Android Automated Input Generation\\
\thanks{This work was supported, in part, by Science Foundation Ireland grant 13/RC/2094\_P2 and co-funded under the European Regional Development Fund through the Southern \& Eastern Regional Operational Programme to Lero - the Science Foundation Ireland Research Centre for Software (\url{www.lero.ie})}
}

\author{
\IEEEauthorblockN{Jordan Doyle, Thomas Laurent, and Anthony Ventresque}
\IEEEauthorblockA{\textit{SFI Lero \&}, \textit{School of Computer Science and Statistics}, \textit{Trinity College Dublin}, Dublin, Ireland\\ doylej51@tcd.ie, tlaurent@tcd.ie, anthony.ventresque@tcd.ie\\ \orcidlink{0009-0001-3448-351X} 0009-0001-3448-351X, \orcidlink{0000-0002-0953-774X} 0000-0002-0953-774X, \orcidlink{0000-0003-2064-1238} 0000-0003-2064-1238}
}

\maketitle

\begin{abstract}\glsunsetall
Android automated test input generation has been a highly researched topic for over a decade and has shown promising results with a variety of approaches. Random input generation is commonly used and the easiest to maintain, but ultimately inefficient. Systematic and search-based approaches produce effective tests but require a disproportionally large generation runtime. Model-based approaches have the additional overhead of modelling the application under test (AUT) but they result in a faster test generation.

In this paper we present Precise AnDRoid Automated Input Generation (\padraig), a model-based test input generation framework that uses a detailed control flow model of the AUT to generate tests that can achieve higher line coverage, with a lower test generation runtime than the state of the art. We compare the line coverage achieved, and the generation runtime of \padraig against 3 state of the art tools, each of which uses a different test input generation technique. Our results, using 19 randomly selected Android apps from the F-Droid application store, show that \padraig achieves, on average, 16\% more coverage of the AUT than the state of the art and it can generate tests with, on average, 84\% less runtime.
\end{abstract}\glsresetall


\section{Introduction}\label{section:introduction}

Mobile applications can be seen in all aspects of modern life. In 2022, the global smartphone penetration rate was estimated at 68\% of a global population of around 7.4 billion~\cite{global2023gsma}. With this massive user base comes a large number of available apps and services with over 3 million apps, representing a revenue of approximately 10.4 billion USD in the Google Play Store alone~\cite{worldwide2022statista}. With their increase in popularity and economic growth it is particularly important to ensure that mobile applications are adequately tested to ensure they are reliable. For instance, a sudden loss of data, or an inconsistent user experience can result in negative feedback and reviews which, coupled with the wide range of applications available, can result in a loss of users and therefore revenue~\cite{roma2013empirical}.

Despite years of research~\cite{joorabchi2013real, amalfitano2013testing, anand2013orchestrated, kochhar2015understanding, amalfitano2017general, linares2017developers, linares2017continuous, kong2018automated, wang2018empirical, pecorelli2022software, samir2022survey}, Android testing is still largely performed manually. Automated test input generation presents a promising approach to alleviating that manual effort and has shown encouraging results in past research using different approaches and achieving varying levels of success. Random~\cite{developers2012ui, anand2012automated, machiry2013dynodroid, muangsiri2017random, paydar2020automated} test input generation is a commonly used and easy to maintain technique but is ultimately inefficient, resulting in a large, ineffective tests. Systematic~\cite{amalfitano2012using, azim2013targeted, moran2017crashscope} and search-based~\cite{mahmood2014evodroid, amalfitano2015agrippin, mao2016sapienz, mariani2021evolutionary, gereziher2023search} techniques show promising results but require an excessive generation runtime. Model-based~\cite{choi2013guided, amalfitano2014mobiguitar, su2017guided, cao2018crawldroid} test input generation is a widely used technique in research. Understanding and modelling the \gls{aut} does create additional initial overhead, but results in a faster test generation and a more effective tests~\cite{samir2022survey}.

This paper introduces \gls{padraig}, a model-based framework for automated test input generation that uses a detailed control flow model of an Android application, to generate effective test inputs that achieve high coverage with a fast test generation runtime. \gls{padraig} is compared against 3 state of the art tools, Monkey~\cite{developers2012ui}, \gls{smog}~\cite{gereziher2023search}, and \gls{stoat}~\cite{su2017guided}, each of which uses a random, search- and model-based test input generation technique, respectively. Our comparison of \gls{padraig} with the state of the art is used to explore the following research questions: 
 
\begin{enumerate}[label=RQ\arabic*,leftmargin=*,align=left]
    \item Can \acrshort{padraig} generate tests that cover more of the application than state of the art tools that use a variety of generation techniques?
    \item How long do the search- and model-based techniques take to generate tests?
\end{enumerate}

To study these questions, Monkey~\cite{developers2012ui}, \gls{smog}~\cite{gereziher2023search}, \gls{stoat}~\cite{su2017guided}, and our proposed approach \gls{padraig}, are applied to 19 randomly selected diverse apps. A comparison of \gls{padraig} and the state of the art shows that \gls{padraig} covers, on average, 16\% more of the \gls{aut} and demonstrates a statistically \textit{large} (see Section~\ref{section:stat_analysis}) improvement over the state of the art. \gls{padraig} is also deterministic and generates tests with an average of 84\% less runtime compared with alternative search- and model-based approaches, while still increasing line coverage. \gls{padraig} as well as the experimental setup and results underlying these conclusions are made available, and can be reproduced using the code and Docker images provided~\cite{doyle2024supplementary}.

The remainder of this paper is structured as follows: Section~\ref{section:ecfg} introduces the detailed control flow model used to guide \glspl{padraig} test generation and section~\ref{section:implementation} outlines the implementation of our proposed approach. Section~\ref{section:experiments} details the experiments supporting the exploration of the above mentioned research questions. Section~\ref{section:results} describes the results of our experiments and discusses these results. Section~\ref{section:related_work} provides a summary of related work in the area of Android test input generation. Finally, Section~\ref{section:threats} outlines possible threats to the validity of this research and Section~\ref{section:conclusion} concludes the paper.
\section{Extended Control Flow Model}\label{section:ecfg}

\gls{padraig} uses an \gls{ecfg} model created by DroidGraph and presented in our previous research~\cite{doyle2023modelling}, to generate precise test inputs. This section first defines the model, including its structure, and then briefly describes how DroidGraph generates this model from an application using both static and dynamic analysis.

\subsection{Model definition}

\begin{definition}[Extended Control Flow Graph]
An \emph{\gls{ecfg}} of a program $P$ with a user interface $U$ is a directed graph $G=(V,E)$ where $V=\{v_1,$ $\ldots,$ $v_n\}$, $n\in \mathbb{N^*}$, is a set of vertices where $v_i$ represents a program point $p_i \in P$ or an interaction $u_i \in U$ and $E=\{e_1, \ldots, e_m\}$, $m\in \mathbb{N^*}$, is a set of arcs (directed edges), where $e_i=(v_j,v_k)$ with $(v_j,v_k) \in V^2$ and $e_i$ represents the flow of control from either, $p_j$ to $p_k$ under some execution of the program $P$ or $u_j$ to $p_k$ after some user interaction.
\end{definition}

The \acrlong{ecfg} $G$ is composed of three types of vertices: $V=V_s\cup V_m\cup V_{\mathit{ui}}$, that represent three elements of a mobile application:
\begin{itemize}
\item $V_s$ is a set of statement vertices. While testing, the focus should be on the code written for the application rather than system or library code. Therefore, such statements are excluded from $V_s$. Statements play an important role by revealing application logic, internal method calls, and the detailed control flow of the application.
\item $V_m$ is a set of method vertices. Similarly, it only comprises methods developed for the application. The method vertices represent the entry points to a method and lead to the statement vertices within. These vertices can be further split into three types: Android lifecycle methods, user input callback methods, and standard Java methods.
\item $V_{\mathit{ui}}$ is a set of interface vertices, representing a \gls{ui} element a user can interact with and providing input to the application.
\end{itemize}

\begin{figure}[tb]
\includegraphics[width=\linewidth]{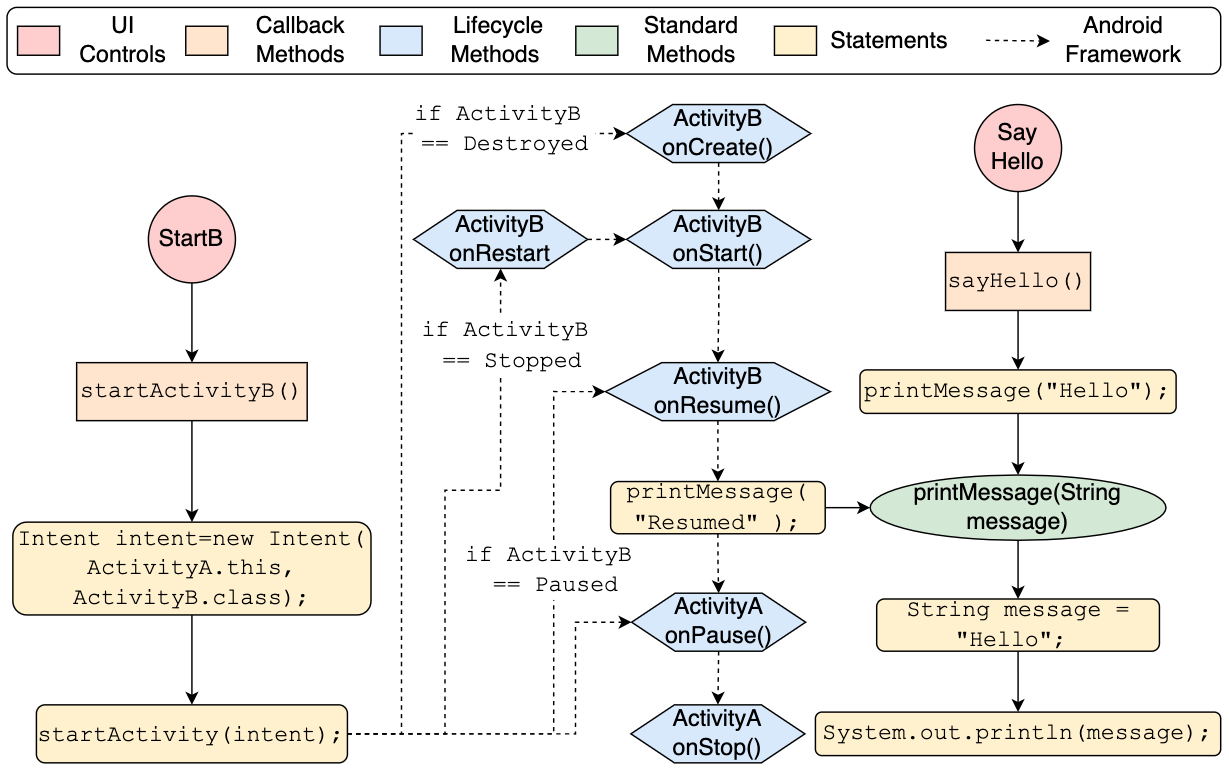}
\caption{A subset of an \acrshort{ecfg}, representing an Android application}
\label{fig:graph_structure}
\end{figure}

Figure~\ref{fig:graph_structure} shows an example subset of the \gls{ecfg} representing an Android application: its internal structure (methods and statements) and the user interactions it contains. This example contains two \gls{ui} controls, 'StartB' and 'SayHello', connected to their listener callback methods \texttt{startActivityB()} and \texttt{sayHello()} respectively. The statements within these methods are linked by edges in their execution order. Method call edges are also shown, for instance the statement \texttt{startActivity(intent)} calls the subsequent lifecycle methods. Some of the edges are shown as a dashed arrow. These edges are not contained in the graph because they originate from outside the application code and the edge used at runtime is determined by the Android framework. It decides which edge to follow based on the current state of the system or application, specifically the called activity. This causes these lifecycle methods to appear as disconnected clusters in the graph. The control flow to and from these clusters is determined at runtime. In order to simplify the example, only statement vertices for the lifecycle method \texttt{ActivityB: onResume()} are included.

In comparison to a traditional \gls{cfg}, the defined \gls{ecfg} introduces \gls{ui} controls, and the control flow to a linked method callback, providing the functional outcome of a user interaction.

\subsection{Model Implementation}

As defined above, the \gls{ecfg} is composed of three types of vertices, that represent three elements of a mobile application. Each vertex type as well as their connected edges are collected using static analysis, dynamic analysis or a combination of both. The static analysis is performed using Soot - a Java optimisation framework, used to analyse, instrument, optimise and visualise Java and Android applications~\cite{lam2011soot}, FlowDroid - a fully context-, field-, object- and flow-sensitive taint analysis tool for Android applications~\cite{arzt2014flowdroid}, and AndroGuard - a python-based tool used for reverse engineering Android apps~\cite{desnos2011android, desnos2012documentation}. The dynamic analysis uses Appium - an open-source project designed to facilitate automation of \ac{ui} interactions with web and mobile platforms~\cite{open2012appium, verma2017mobile}, to perform an enhanced \gls{dfs} on the application \gls{ui} using an instrumented version of the \gls{apk}. The vertices and edges of the \gls{ecfg} are populated as follows:

\begin{itemize}
\item \textbf{Statement} vertices are created based on FlowDroid \texttt{UnitGraph} objects created for each method in the AUT. Each \texttt{UnitGraph} contains a unit chain detailing all the statements and their execution order within the method. Intra-procedural call edges are determined using Soot's inter- and intra- procedural analysis.
\item \textbf{Method} vertices are created by retrieving all the methods, for each class, identified by FlowDroid in the AUT. The methods and the inter-procedural calls between them are also gathered from the call-graph generated by AndroGuard. The model is limited to internal components by filtering external library methods such as AndroidX. Before adding a method to the model, it is categorised as an activity lifecycle method, input listener method, or a standard Java method.
\item \textbf{Interface} controls are identified in FlowDroid by parsing the Android XML layout files included in the compiled \gls{apk}. FlowDroid is capable of identifying controls and their XML attributes when declared in XML layouts but controls declared in the Java source code are identified and added to the model during the dynamic analysis. The edge between \gls{ui} controls and their associated listener callback methods are populated during the dynamic analysis, the instrumented \gls{apk} provides a log of method calls and the \gls{ui} interaction that initiated them.
\end{itemize}

As mentioned, the \gls{ecfg} and DroidGraph are presented in our previous research~\cite{doyle2023modelling} where we explore methods of modelling application code structures and the benefits of combining both static and dynamic analysis. This paper presents \gls{padraig}, a test input generator that uses DroidGraph to create an \gls{ecfg} model and uses this model to generate strong tests in an efficient way.
\section{\padraig Implementation}\label{section:implementation}

\begin{figure}[tb]
\centering
\includegraphics[width=\linewidth]{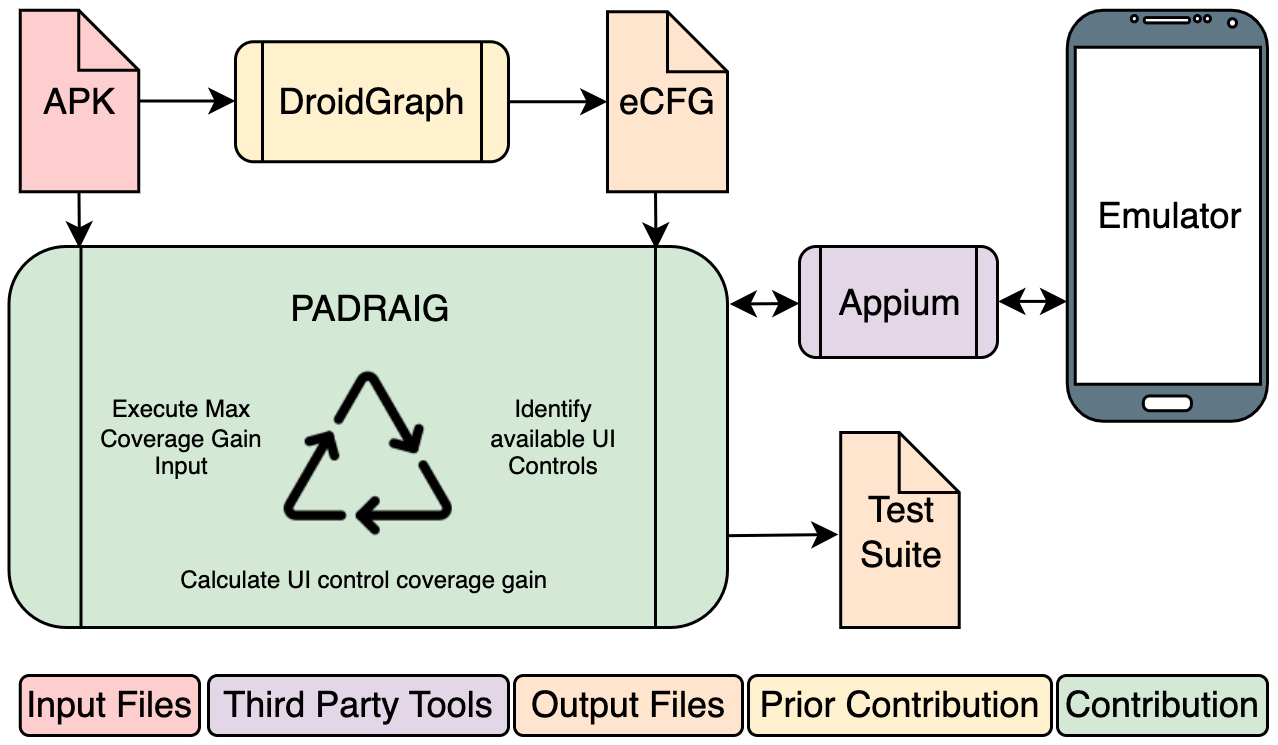}
\caption{The component structure of the \acrshort{padraig} framework}
\label{fig:padraig_overview}
\end{figure}

\gls{padraig} is composed of various tools and components that enable it to perform model-based test input generation for an Android application. Figure~\ref{fig:padraig_overview} provides an overview of all the processes involved and how they interact. The \gls{aut} is provided as an input, in the form of an \gls{apk} file, to both DroidGraph and \gls{padraig}. DroidGraph first creates an \gls{ecfg} of the \gls{aut}, as discussed in Section~\ref{section:ecfg}. \gls{padraig} explores the \gls{aut} and generates inputs simultaneously using Appium~\cite{open2012appium, verma2017mobile} and the \gls{ecfg} created by DroidGraph. Appium is highly integrated with Android's UI-Automator which provides detailed knowledge of \gls{ui} controls currently available to interact with, and their locations on the screen. Using the \gls{ecfg}, the \gls{apk}, and Appium coupled with an Android emulator, \gls{padraig} generates and outputs a test for the \gls{aut}, consisting of a sequence of inputs. The inputs that \gls{padraig} generates are currently limited to taps, while inputs such as swipes and text inputs are planned for future work.

Algorithm~\ref{algo:padraig} shows how \gls{padraig} generates test inputs, given a budget chosen by the user in the form of $N$, the number of interactions a test should contain. The algorithm starts in the default launch activity of the \gls{aut} by launching the app on the Android emulator. Using Appium, it retrieves the \gls{ui} controls available for the currently displayed activity, including the back button \gls{ui} control (line~\ref{line:available_controls} in algorithm~\ref{algo:padraig}). Each of the available \gls{ui} controls is assigned a fitness value based on the potential coverage gain it provides (line~\ref{line:coverage_gain} in algorithm~\ref{algo:padraig}). Algorithm~\ref{algo:fitness} shows how the potential coverage gain of a \gls{ui} control is calculated. \gls{padraig} determines the next input by choosing the \gls{ui} control with the maximum fitness value, i.e. the maximum potential coverage gain. It is possible that multiple \gls{ui} controls are assigned the same fitness value. In this scenario, \gls{padraig} uses a frequency strategy that chooses the least used UI control from the inputs that have equal fitness values (line~\ref{line:least_frequently} in algorithm~\ref{algo:padraig}). The inputs already performed on the \gls{aut} are tracked so \gls{padraig} can determine which \gls{ui} control has been used the least (line~\ref{line:tracker} in algorithm~\ref{algo:padraig}). When \gls{padraig} decides to interact with a \gls{ui} control, the associated \gls{ui} control vertex, as well as the vertices in the path starting from it, are marked as visited in the \gls{ecfg} (line~\ref{line:visit_trace} in algorithm~\ref{algo:padraig}). Visited vertices reduce the potential coverage gain of a \gls{ui} control and therefore the fitness of repeated inputs. This allows for repeated inputs when needed but does not allow excessive repetition. 

\begin{algorithm}[ht]
    \caption{\acrshort{padraig} generating a sequence of inputs}
    \label{algo:padraig}
    \begin{algorithmic}[1]
    \Require $graph$, the \acrshort{ecfg} of the \acrshort{aut}
    \Require $N$, the number of inputs to generate.
    \Ensure $test$, a test, i.e, a sequence of inputs.
    \State $test \gets \emptyset$
    \State $tracker \gets \emptyset$

    \State launchApp()

    \For{$count \gets 0$ to $N$}
        \State $controls \gets$ getAvailableControls() \label{line:available_controls}
        \State $covGains \gets \emptyset$
        
        \For{$c \in controls$}
            \State $g \gets graph.$calculateCoverageGain($c$) \label{line:coverage_gain}
            \State $covGains.$add($c$, $g$)
        \EndFor

        \State $maxControls \gets$ maxValue($covGains$)
        \If{$maxControls.$size() $> 1$}
            \State $next \gets tracker.$leastFrequent($maxControls$) \label{line:least_frequently}
        \Else 
            \State $next \gets maxControls.$get($0$)
        \EndIf 

        \State $next.$click()
        \State $tracker.$addInput($next$) \label{line:tracker}
        \State $test.$add($next.$getAdbCommand())
        \State $graph.$markCoveredVertices($next$) \label{line:visit_trace}
    \EndFor

    \State \Return $test$
    \end{algorithmic}
\end{algorithm}

As mentioned, the fitness value of the \gls{ui} controls are determined by the potential coverage gain they provide. Algorithm~\ref{algo:fitness} shows how the fitness value for a \gls{ui} control vertex \textit{v} is calculated. \gls{padraig} traverses the paths starting at this vertex, updating the fitness value based on the vertices it encounters, until the paths end. The returned value is a count of the number of vertices that will potentially be visited as a result of the input, but have not already been visited previously. Additionally vertices demonstrating application features, such as the launch of a new activity, increase the fitness value of the \gls{ui} control, as they indicate potentially unexplored territory in the \gls{aut}. The calculated coverage gain is an overestimate of the coverage gain that will be achieved by using the \gls{ui} control. For example, when an if-else statement is encountered, both branches are included in the potential coverage gain, despite only one branch executing each time the \gls{ui} control is used. A more accurate coverage value would require a data flow analysis which is supported by FlowDroid, the primary static analysis tool used to generate the \gls{ecfg}, but is not included in \gls{padraig}.

\begin{algorithm}[ht]
    \caption{Calculating the potential coverage gain of a \acrshort{ui} control vertex in an \acrshort{ecfg}}
    \label{algo:fitness}
    \begin{algorithmic}[1]
    \Require $v$, a \acrshort{ui} control vertex in the app \acrshort{ecfg}
    \Ensure $g$, potential coverage gain of the \acrshort{ui} control vertex $v$
    \Procedure{calculateCoverageGain}{$v$}
        \If{$v.$hasVisit()}
            \State $g \gets 0$
        \Else 
            \State $g \gets 1$
        \EndIf 
        
        \For{$e \in v$.getOutGoingEdges()}
            \State $t \gets e.$getEdgeTarget()  
    	\State $g \gets g +$ calculateCoverageGain($t$)
        \EndFor
    
        \If{$v.$getAttribute$(type) == STATEMENT$ \textbf{and} $v.$hasFeature()}
            \State $g \gets g + 1$
        \EndIf
        
        \State \Return $g$
    \EndProcedure
    \end{algorithmic}
\end{algorithm}
\section{Experiments}\label{section:experiments}

This section describes the experiments conducted to assess the test input generation capabilities of \gls{padraig} compared to the state of the art. First, it outlines the research questions explored. Then, it introduces the applications used as benchmarks. Then, it describes the protocols and metrics used to investigate the research questions. Finally, it discusses a statistical analysis performed on the experiment results. The experimental setup and results are made available online~\cite{doyle2024supplementary}.

\subsection{Research Questions}\label{section:rqs}

In order to understand the contribution of \gls{padraig}, the following research questions are explored:
\begin{enumerate}[label=RQ\arabic*,leftmargin=*,align=left]
    \item \textbf{Can \acrshort{padraig} generate tests that cover more of the application than state of the art tools that use a variety of generation techniques?}

    This research question compares the line coverage achieved by \gls{padraig} and 3 state of the art automated test input generation frameworks, Monkey~\cite{developers2012ui}, \gls{smog}~\cite{gereziher2023search}, and \gls{stoat}~\cite{su2017guided}. It explores how much of the applications' code a test produced by each tool execute, reflecting how complete a test suite the tools can create.
    
    \item \textbf{How long do the search- and model-based techniques take to generate a test?}

    This research question explores the runtime required by search- and model-based test input generation techniques to produce a test that can achieve suitable coverage of the \gls{aut} and demonstrates the feasibility of the approach being used in practice. The runtime of model-based test input generation techniques also include the time taken to generate a model of the \gls{aut} in order to encompass the entire test generation process. Note: Monkey is not included in order to maintain a fair comparison, its random nature makes the test generation artificially fast, i.e. it does not generate tests using a ``smart'' method.
\end{enumerate}

\subsection{Subject applications}

In order to study the above research questions, each test input generation technique was applied to a diverse set of Android applications. A random selection of applications were chosen from the F-Droid market place~\cite{limited2010fdroid}, which contains over 4000 apps. In order to represent the different types of applications developed in practice, one application was randomly selected from each of the 20 categories (excluding Games) of applications found in F-Droid. Before the selection was made, unsuitable apps were filtered out based on 3 criteria: 

\begin{enumerate}
  \item apps in the games category, as they often include game development frameworks such as unity, which are not supported by \gls{padraig}. 
  \item apps with an Android \gls{sdk} version less than 16 (too old) or greater than 29 (not yet supported by FlowDroid). 
  \item apps that have not been maintained, i.e., not updated within the last 10 years.
\end{enumerate}

Three apps used in previous research~\cite{doyle2021improving} are also included to demonstrate that previous results traversing the \gls{ecfg} are consistent with the approach used by \gls{padraig}. Table~\ref{tab:subject_apps} shows the list of apps used in the experiments. The \gls{loc} for each app is obtained using the Statistic~\cite{statistic} plugin in Android Studio. All corresponding \gls{apk} files are made available in~\cite{doyle2024supplementary}.

\begin{table}[ht]
    \centering
    \setlength{\tabcolsep}{2pt}
    \caption{Applications used to compare \acrshort{padraig} with the state of the art}
    \label{tab:subject_apps}
    \begin{tabular}{@{}llrr@{}}
        \toprule
        \textbf{App name} & \textbf{Category} & \textbf{Version} & \textbf{LoC}\\
        \midrule
        Activity Lifecycle & Science \& Education & 1 & 909 \\
        Ad-Free & Internet & 41 & 5,608 \\
        Battery Live & Theming & 13 & 1,062 \\
        Camera Roll & Multimedia & 36 & 21,006 \\
        Contact Book & Phone \& SMS & 1 & 2,904 \\
        Drinks & Reading & 32 & 1,820 \\
        Git Quick Reference & Development & 7 & 2,832 \\
        GPSTest & Navigation & 18093 & 134,547 \\
        Loyalty Card Keychain & Money & 39 & 5,998 \\
        MoClock & Time & 4 & 499 \\
        PIN Mnemonic & Security & 6 & 2,304 \\ 
        Pixel Filter & Graphics & 24 & 1,435 \\
        Simple Explorer & System & 67 & 13,426 \\
        Simple Todo & Time & 5 & 3,216 \\
        Taskkeeper & Writing & 6 & 1,626 \\
        Timetable & Science \& Education & 17 & 8,477 \\
        Volume Control & System & 32 & 2,184 \\
        Webradio & Connectivity &5 & 1,709 \\
        Wine Cellar & Sports \& Health & 4 & 3,804 \\
        \bottomrule
    \end{tabular}
\end{table}

\subsection{Experimental Procedure}

In order to investigate the above research questions, each of the frameworks, Monkey~\cite{developers2012ui}, \gls{smog}~\cite{gereziher2023search}, \gls{stoat}~\cite{su2017guided}, and \gls{padraig} are applied to all the applications in Table~\ref{tab:subject_apps}. In order to account for randomness, Monkey, \gls{smog}, and \gls{stoat} are executed 10 times for each application. \gls{padraig} uses a deterministic algorithm so it does not contain randomness, and therefore is executed once per application. The maximum number of inputs per test in~\cite{gereziher2023search} and~\cite{su2017guided} was 50 and 30 respectively, so each tool was given a budget of 50 interactions per test for a fair comparison.

Monkey~\cite{developers2012ui} was executed under two settings: limiting it to only click interactions (Monkey Click), and allowing it to use all interactions it supports (Monkey All). Both settings also had a 500 ms threading time and were limited to packages belonging to the \gls{aut}. These settings are applied to reduce stress testing on the \gls{aut} and ensure Monkey generated inputs are only applied to the \gls{aut}. The same settings presented in~\cite{gereziher2023search} were allocated to \gls{smog}, a population of 10 tests and search budget of 10 generations. The \gls{fsm} model construction and test generation that \gls{stoat} performs is time based. As presented in~\cite{su2017guided} \gls{stoat} was given a budget of one hour to generate an \gls{fsm} and another two hours to generate a test.

The overall line coverage achieved by each of the generated tests was measured using ACVTool~\cite{pilgun2020acvtool} which does not use the original source code. The measured line coverage is based on a Smali representation of the bytecode. A comparison of the coverage achieved for each application by Monkey~\cite{developers2012ui}, \gls{smog}~\cite{gereziher2023search}, \gls{stoat}~\cite{su2017guided}, and \gls{padraig} serves to answer RQ1, while comparing the runtime required to generate the tests answers RQ2. All experiments and results can be reproduced using the code and Docker images provided~\cite{doyle2024supplementary}.

\subsection{Statistical Analysis}\label{section:stat_analysis}

In order to demonstrate the significance and level of improvement provided by \gls{padraig} over Monkey, \gls{smog}, and \gls{stoat}, each approach is compared using the Mann-Whitney U test, a non parametric test telling whether there is a significant difference among the results. The value $\alpha = 0.05$ is used as a confidence value of the null hypothesis that there is no significant difference. In case of significant difference, Vargha and Delaney’s \Atwelve effect size is used to assess the strength of the significance; if \Atwelve is greater than 0.5, the results of \gls{padraig} are significantly better than those of Monkey, \gls{smog}, and \gls{stoat}. By following Kitchenham et al.’s classification~\cite{kitchenham2017robust}, the following categories of strength are identified: negligible when $\Atwelve \in (0.5, 0.556)$, small when $\Atwelve \in [0.556, 0.638)$, medium when $\Atwelve \in [0.638, 0.714)$, and large when $\Atwelve \geq 0.714$. Similar categories can be identified for $\Atwelve < 0.5$, i.e., when \gls{padraig} is significantly worse. In this comparison, we count the number of experiments where one approach is better than the other, and with which strength.
\section{Results}\label{section:results}

This section describes the results obtained from the experiments discussed in Section~\ref{section:experiments} and how they relate to the defined research questions.

\subsection{RQ1 Can \padraig generate tests that cover more of the application than state of the art tools that use a variety of generation techniques?} 

\begin{figure}[tb]
    \centering
    \input{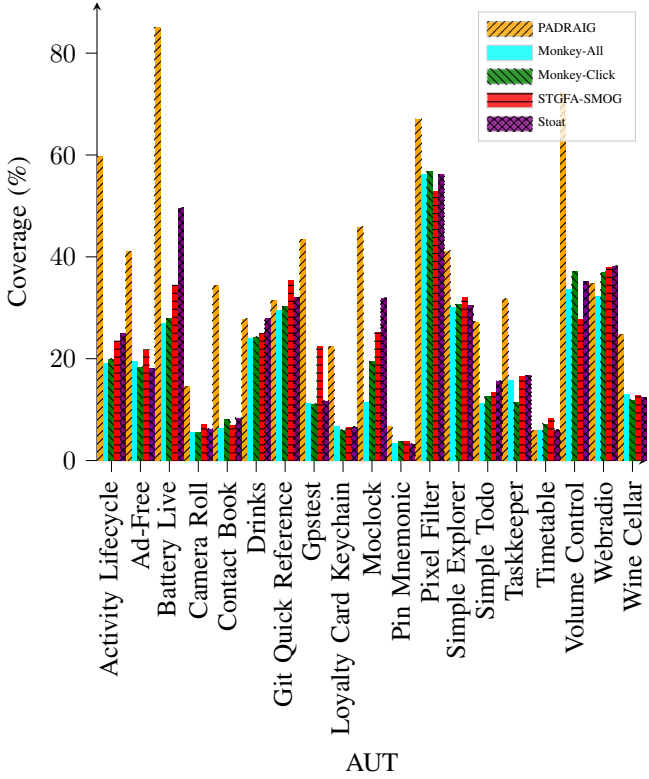}
    \caption{Coverage (\%) achieved by \acrshort{padraig} and the state of the art}
    \label{fig:padraig_line_bar}
\end{figure}

Table~\ref{tab:padraig_line_coverage}, and Figure~\ref{fig:padraig_line_bar} 
show a comparison of the line coverage achieved by each approach. In Table~\ref{tab:padraig_line_coverage}, for each application, the maximum line coverage achieved for that app is shown in bold.

\begin{table*}[ht]
    \centering
    \setlength{\tabcolsep}{2pt}
    \caption{Coverage (\%) achieved by \acrshort{padraig} and the state of the art}
    \label{tab:padraig_line_coverage}
    \begin{tabular}{@{}lrrrrrrrrrrrrr@{}}
    \toprule
        \multirow{2}{*}{\textbf{App Name}} & \multirow{2}{*}{\textbf{\padraig}} & \multicolumn{3}{c}{\textbf{Monkey All}} & \multicolumn{3}{c}{\textbf{Monkey Click}} & \multicolumn{3}{c}{\textbf{STGFA-SMOG}} & \multicolumn{3}{c}{\textbf{Stoat}}\\
        \cmidrule(lr){3-5}\cmidrule(lr){6-8}\cmidrule(lr){9-11}\cmidrule(lr){12-14}
        & & \textbf{Avg} & \textbf{Min} & \textbf{Max} & \textbf{Avg} & \textbf{Min} & \textbf{Max} & \textbf{Avg} & \textbf{Min} & \textbf{Max}  & \textbf{Avg} & \textbf{Min} & \textbf{Max}\\
        \midrule
        Activity Lifecycle & \textbf{59.8} & 19.2 & 11.24 & 28.19 & 19.99 & 13.62 & 24.77 & 23.51 & 12.55 & 28.39 & 25.01 & 21.64 & 41.07 \\
        Ad-Free & \textbf{41.11} & 19.57 & 18.13 & 27.34 & 18.44 & 18.13 & 19.4 & 21.75 & 18.22 & 25.23 & 18.14 & 18.14 & 18.14 \\
        Battery Live & \textbf{85.08} & 26.9 & 25.63 & 28.29 & 27.95 & 25.98 & 30.4 & 34.48 & 25.63 & 78.44 & 49.63 & 26.03 & 63.47 \\
        Camera Roll & \textbf{14.54} & 5.53 & 5.43 & 5.69 & 5.6 & 5.43 & 6.91 & 7.25 & 5.5 & 10.12 & 6.2 & 5.71 & 9.06 \\
        Contact Book & \textbf{34.47} & 6.41 & 6.41 & 6.41 & 8.13 & 6.41 & 11.58 & 6.88 & 6.41 & 8.48 & 8.39 & 6.55 & 13.76 \\
        Drinks & 27.88 & 24.1 & 23.75 & 25.0 & 24.33 & 23.75 & 25.09 & 24.97 & 23.75 & 26.56 & 27.91 & 25.09 & \textbf{29.91} \\
        Git Quick Reference & 31.49 & 29.5 & 22.84 & 32.65 & 30.3 & 28.65 & 33.1 & 35.48 & 29.96 & 36.41 & 32.02 & 28.53 & \textbf{37.66} \\
        Gpstest & \textbf{43.42} & 11.25 & 11.09 & 11.44 & 11.18 & 11.09 & 11.55 & 22.52 & 21.16 & 25.59 & 11.71 & 11.2 & 16.09 \\
        Loyalty Card Keychain & \textbf{22.41} & 6.75 & 5.81 & 7.87 & 6.01 & 5.81 & 6.63 & 6.63 & 5.81 & 7.46 & 6.61 & 6.3 & 7.72 \\
        Moclock & \textbf{45.87} & 11.45 & 6.71 & 21.84 & 19.61 & 6.71 & 35.57 & 25.23 & 18.49 & 34.87 & 31.98 & 29.49 & 44.46 \\
        Pin Mnemonic & \textbf{6.69} & 3.48 & 3.32 & 3.79 & 3.8 & 3.6 & 4.07 & 3.92 & 3.44 & 4.34 & 3.34 & 3.32 & 3.41 \\
        Pixel Filter & \textbf{67.08} & 56.27 & 54.17 & 60.57 & 56.84 & 54.17 & 58.94 & 52.83 & 29.47 & 60.92 & 56.27 & 56.03 & 58.43 \\
        Simple Explorer & \textbf{41.21} & 30.15 & 29.29 & 31.63 & 30.69 & 28.91 & 33.03 & 32.15 & 29.63 & 35.84 & 30.47 & 29.72 & 34.6 \\
        Simple Todo & 27.27 & 11.3 & 11.01 & 11.99 & 12.65 & 11.01 & 18.54 & 13.55 & 11.01 & 18.24 & 15.64 & 12.09 & \textbf{33.31} \\
        Taskkeeper & 31.79 & 15.75 & 10.72 & 18.76 & 11.39 & 10.72 & 16.93 & 16.69 & 10.72 & 40.2 & 16.73 & 10.92 & \textbf{42.73} \\
        Timetable & 5.92 & 6.06 & 5.91 & 7.38 & 7.22 & 5.91 & 9.18 & 8.34 & 6.76 & \textbf{10.97} & 6.04 & 5.87 & 6.88 \\
        Volume Control & \textbf{72.29} & 33.67 & 30.44 & 39.48 & 37.16 & 33.54 & 41.08 & 27.76 & 4.64 & 40.55 & 35.17 & 32.53 & 46.6 \\
        Webradio & 34.83 & 32.36 & 29.4 & 39.52 & 37.04 & 28.37 & 40.81 & 37.95 & 28.77 & 39.76 & 38.25 & 30.09 & \textbf{42.89} \\
        Wine Cellar & \textbf{24.81} & 13.11 & 11.9 & 15.54 & 11.9 & 11.9 & 11.9 & 12.9 & 11.74 & 21.03 & 12.46 & 11.74 & 18.95 \\
        \bottomrule
    \end{tabular}
\end{table*}

The tests generated by \gls{padraig} achieve higher line coverage of the \gls{aut} than the average achieved by Monkey with all interactions, Monkey with only click interactions, \gls{smog}, and \gls{stoat} in 18, 17, 16, and 16 out of 19 apps respectively. While the coverage achieved by the state of the art is comparable in some cases, for example the Drinks app, \gls{padraig} covers, on average 16\%, and up to 50\% more of the \gls{aut} than the state of the art for the majority of the \glspl{aut}. There are cases where \gls{padraig} did not perform as well as the state of the art, for example, Monkey, \gls{smog}, and \gls{stoat} achieved higher line coverage than \gls{padraig} in the Timetable app. There are many factors that could cause this, for instance, the AUT functionality could be gesture orientated or require system events which \gls{padraig} does not currently support. Despite these few instances, the results clearly demonstrate that \gls{padraig} is not only effective at covering the \gls{aut} but better than the state of the art for the majority of the \glspl{aut}.

The line coverage achieved by \gls{padraig} and the state of the art appears to be low in many of the apps. There are a number of explanations for this, for example some applications require content, such as a contact app, where a portion of the interface is only visible after the user has added a contact. Also larger applications may require a larger number of inputs in order to cover a greater portion of the code.

Table~\ref{tab:padraig_coverage_stat_comparison} further demonstrates the level of significant improvement provided by \gls{padraig} over the state of the art. It shows that \glspl{padraig} model-based approach is statistically significantly better than Monkey with all interactions, Monkey with only click interactions, \gls{smog}, and \gls{stoat} in 17, 16, 16, and 16 out of 19 apps respectively.

\begin{table}[ht]
    \centering
    \footnotesize
    \setlength{\tabcolsep}{2pt}
    \caption{Statistical significance of the coverage improvements between \acrshort{padraig} and the state of the art}
    \label{tab:padraig_coverage_stat_comparison}
    \begin{tabular}{@{}lcccc@{}}
    \toprule
        \multirow{2}{*}{\textbf{App Name}} & \multicolumn{2}{c}{\textbf{Monkey}} & \multirow{2}{*}{\textbf{STGFA-SMOG}} & \multirow{2}{*}{\textbf{Stoat}}\\
        \cmidrule(lr){2-3}
        & \textbf{All Inputs} & \textbf{Click Inputs} & & \\
        \midrule
        Activity Lifecycle & \betterLarge & \betterLarge & \betterLarge & \betterLarge \\
        Ad-Free & \betterLarge & \betterLarge & \betterLarge & \betterLarge \\
        Battery Live & \betterLarge & \betterLarge & \betterLarge & \betterLarge \\
        Camera Roll & \betterLarge & \betterLarge & \betterLarge & \betterLarge \\
        Contact Book & \betterLarge & \betterLarge & \betterLarge & \betterLarge \\
        Drinks & \betterLarge & \betterLarge & \betterLarge & \worseLarge \\
        Git Quick Reference & \same & \same & \worseLarge & \same \\
        Gpstest & \betterLarge & \betterLarge & \betterLarge & \betterLarge \\
        Loyalty Card Keychain & \betterLarge & \betterLarge & \betterLarge & \betterLarge \\
        Moclock & \betterLarge & \betterLarge & \betterLarge & \betterLarge \\
        Pin Mnemonic & \betterLarge & \betterLarge & \betterLarge & \betterLarge \\
        Pixel Filter & \betterLarge & \betterLarge & \betterLarge & \betterLarge \\
        Simple Explorer & \betterLarge & \betterLarge & \betterLarge & \betterLarge \\
        Simple Todo & \betterLarge & \betterLarge & \betterLarge & \betterLarge \\
        Taskkeeper & \betterLarge & \betterLarge & \betterLarge & \betterLarge \\
        Timetable & \betterLarge & \worseLarge & \worseLarge & \betterLarge \\
        Volume Control & \betterLarge & \betterLarge & \betterLarge & \betterLarge \\
        Webradio & \same & \same & \worseLarge & \worseLarge \\
        Wine Cellar & \betterLarge & \betterLarge & \betterLarge & \betterLarge \\
        \bottomrule
    \end{tabular}
    
    \same: no statistically significant difference between the approaches. \ding{51}: \acrshort{padraig} tests are better than the approach on top for the metric, \ding{55} means that it is worse; the num. of symbols is the strength: negligible (\betterNegl, \worseNegl), small (\betterSmall, \worseSmall), medium (\betterMedium, \worseMedium), large (\betterLarge, \worseLarge)
\end{table}

\subsection{RQ2 How long do the search- and model-based techniques take to generate a test?}

The runtime for creating a test for an application is not commonly evaluated in past research. However, it is arguably a compelling factor in the adoption of an automated test input generation technique. Automated test input generation is intended to remove the slow, labour intensive process of creating tests manually, and therefore needs to be fast and efficient. The faster tests are created and executed on the \gls{aut}, the faster faults can be detected and fixed. Efficient generation techniques can lead to more extensive and comprehensive test suites that can keep up with the fast pace of the mobile industry.

While Monkey is a test input generator, the algorithm used to determine suitable inputs and its computational complexity can be compared with that of a simple random number generator. As such, Monkey's test generation runtime is artificially fast and will always be several orders of magnitude better than any alternative approach, regardless of the line coverage they achieve. Therefore, when comparing the runtime of \gls{padraig} with the state of the art, Monkey is not included in order to maintain a fair comparison. 

Table~\ref{tab:padraig_runtime} shows a comparison of the runtime required by \gls{smog}, \gls{stoat}, and \gls{padraig} to generate a test for each \gls{aut}. The runtime shown for \gls{padraig} and \gls{stoat} includes the time taken to generate a model of the \gls{aut}, and the minimum runtime required for each app is shown in bold.

\begin{table*}[ht]
    \centering
    \setlength{\tabcolsep}{2pt}
    \caption{Runtime (in minutes) of search- and model-based techniques}
    \label{tab:padraig_runtime}
    \begin{tabular*}{\textwidth-20pt}{@{\extracolsep{\fill}}lrrrrrrr@{}}
    \toprule
        \multirow{2}{*}{\textbf{App Name}} & \multirow{2}{*}{\textbf{\padraig}} & \multicolumn{3}{c}{\textbf{SMOG}} & \multicolumn{3}{c}{\textbf{Stoat}}\\
        \cmidrule(lr){3-5}\cmidrule(lr){6-8}
        & & \textbf{Avg} & \textbf{Min} & \textbf{Max} & \textbf{Avg} & \textbf{Min} & \textbf{Max}\\
        \midrule
        Activity Lifecycle & \textbf{17.7} & 156.59 & 142.88 & 187.33 & 192.72 & 192.52 & 192.78 \\
        Ad-Free & \textbf{17.85} & 185.12 & 167.17 & 207.03 & 189.71 & 163.52 & 192.63 \\
        Battery Live & \textbf{18.1} & 146.09 & 126.57 & 169.33 & 192.6 & 192.45 & 192.62 \\
        Camera Roll & \textbf{22.3} & 159.52 & 137.88 & 185.27 & 192.6 & 192.6 & 192.62 \\
        Contact Book & \textbf{17.9} & 142.66 & 133.73 & 154.95 & 191.95 & 190.35 & 192.6 \\
        Drinks & \textbf{22.65} & 150.88 & 143.47 & 162.22 & 192.6 & 192.47 & 192.62 \\
        Git Quick Reference & \textbf{23.6} & 146.45 & 131.43 & 169.85 & 192.59 & 192.45 & 192.62 \\
        Gpstest & \textbf{48.12} & 168.97 & 143.0 & 226.38 & 192.61 & 192.58 & 192.62 \\
        Loyalty Card Keychain & \textbf{22.45} & 154.34 & 144.88 & 185.88 & 192.68 & 192.47 & 192.78 \\
        Moclock & \textbf{19.88} & 138.62 & 124.58 & 162.88 & 192.61 & 192.6 & 192.62 \\
        Pin Mnemonic & \textbf{18.87} & 139.89 & 130.58 & 165.38 & 192.61 & 192.58 & 192.62 \\
        Pixel Filter & \textbf{65.43} & 155.47 & 132.0 & 190.12 & 192.58 & 192.57 & 192.58 \\
        Simple Explorer & \textbf{21.52} & 147.12 & 127.03 & 174.28 & 192.59 & 192.58 & 192.6 \\
        Simple Todo & \textbf{19.45} & 140.6 & 123.88 & 165.65 & 192.06 & 191.7 & 192.62 \\
        Taskkeeper & \textbf{16.82} & 134.19 & 119.57 & 158.25 & 186.86 & 185.95 & 188.43 \\
        Timetable & \textbf{31.58} & 148.57 & 139.28 & 156.7 & 192.59 & 192.47 & 192.62 \\
        Volume Control & \textbf{57.68} & 150.46 & 137.63 & 181.58 & 192.59 & 192.48 & 192.6 \\
        Webradio & \textbf{19.8} & 132.59 & 125.53 & 142.72 & 192.58 & 192.45 & 192.6 \\
        Wine Cellar & \textbf{20.77} & 149.31 & 133.05 & 175.73 & 192.6 & 192.6 & 192.62 \\
        \bottomrule
    \end{tabular*}
\end{table*}

The runtime required by \gls{padraig} is on average 2.4 hours or 84\% less less than the runtime required by \gls{smog}, a search based technique, and \gls{stoat}, an alternative model-based technique. Due to \gls{stoat} being time based, an argument can be made for lowering the allowed time for \gls{stoat} to generate an \gls{fsm} and test inputs to achieve the same speed as \gls{padraig}. However, as seen in Table~\ref{tab:padraig_line_coverage} and Figure~\ref{fig:padraig_line_bar}, \gls{stoat} fails to generate tests with better line coverage of the \gls{aut} than \gls{padraig} despite its current time budget, which is much larger than \gls{padraig}'s. This suggests that, given a time budget equal to \gls{padraig}'s runtime, \gls{stoat} would perform even worse. Search-based techniques, such as \gls{smog}, are computationally expensive, they require many tests to be created and run before returning the best test in the population, this therefore increases the runtime required. 

While the scalability of each approach was not evaluated in this paper, the runtime achieved by \gls{padraig} appears unaffected by the size of the \gls{aut}, for example, \glspl{padraig} runtime using the app Simple Explorer with 13,426 \gls{loc} is 21.52 minutes. \gls{padraig} is also deterministic, which means that it can generate effective tests with a lower runtime, while providing the same level of success for every test it generates.

Table~\ref{tab:padraig_runtime_stat_comparison} also illustrates the level of significant improvement provided by \gls{padraig} over the state of the art. It shows that the test generation runtime of \glspl{padraig} model-based approach is statistically significantly better than the state of the art in all the \glspl{aut}.

\begin{table}[ht]
    \centering
    \setlength{\tabcolsep}{2pt}
    \caption{Statistical significance of the runtime improvements between \acrshort{padraig} and the state of the art}
    \label{tab:padraig_runtime_stat_comparison}
    \begin{tabular*}{\columnwidth-20pt}{@{\extracolsep{\fill}}lcc@{}}
    \toprule
        \textbf{App Name} & \textbf{STGFA-SMOG} & \textbf{Stoat}\\
        \midrule
        Activity Lifecycle & \betterLarge & \betterLarge \\
        Ad-Free & \betterLarge & \betterLarge \\
        Battery Live & \betterLarge & \betterLarge \\
        Camera Roll & \betterLarge & \betterLarge \\
        Contact Book & \betterLarge & \betterLarge \\
        Drinks & \betterLarge & \betterLarge \\
        Git Quick Reference & \betterLarge & \betterLarge \\
        Gpstest & \betterLarge & \betterLarge \\
        Loyalty Card Keychain & \betterLarge & \betterLarge \\
        Moclock & \betterLarge & \betterLarge \\
        Pin Mnemonic & \betterLarge & \betterLarge \\
        Pixel Filter & \betterLarge & \betterLarge \\
        Simple Explorer & \betterLarge & \betterLarge \\
        Simple Todo & \betterLarge & \betterLarge \\
        Taskkeeper & \betterLarge & \betterLarge \\
        Timetable & \betterLarge & \betterLarge \\
        Volume Control & \betterLarge & \betterLarge \\
        Webradio & \betterLarge & \betterLarge \\
        Wine Cellar & \betterLarge & \betterLarge \\
        \bottomrule
    \end{tabular*}
    
    \same: no statistically significant difference between the approaches. \ding{51}: \acrshort{padraig} tests are better than the approach on top for the metric, \ding{55} means that it is worse; the num. of symbols is the strength: negligible (\betterNegl, \worseNegl), small (\betterSmall, \worseSmall), medium (\betterMedium, \worseMedium), large (\betterLarge, \worseLarge)
\end{table}
\section{Related work}\label{section:related_work}

This section outlines previous work in the area of automated test input generation for Android applications. This area has grown to such an extent over the last decade that it can be further classified into subcategories based on the methods used. The most used methods are random, systematic, search-, and model-based input generation.

\subsection{Random}

Using a random approach for generating test inputs is the simplest method but it is generally very inefficient and leads to a high number of redundant (provide no contribution to the test) and repetitive (lower the diversity of the test) interactions. An example of this is Monkey~\cite{developers2012ui}, a program that runs on an emulator or device and generates pseudo-random streams of user events such as clicks, touches, or gestures, as well as a number of system-level events. In practice, Monkey does not support keyboard input or system notifications, and most of the random inputs applied to the interface do nothing within the application (e.g., tapping an area of the screen with no actionable \acrshort{ui} element). Dynodroid~\cite{machiry2013dynodroid}, another well-known random input generator, tries to overcome this issue by applying a Frequency Strategy and Biased-Random Strategy, to increase the efficiency of the inputs it selects. The frequency strategy uses the inputs that have been least frequently used in the past and the biased-random strategy uses the inputs that are relevant in most contexts. Unlike Monkey, Dynodroid also supports keyboard input and system notifications, however, this is possibly its downfall. For instance, in order to provide system events (e.g, keyboard input, notifications, changing device orientation), Dynodroid needs to instrument the Android framework and in doing this it became hard to update and maintain, which has led to the framework becoming outdated and unused. More recently, MonkeyImprover~\cite{paydar2020automated} proposed refactoring the \gls{gui} of the \gls{aut} to make \gls{ui} controls that lead to more complex background code more prominent, and thus increasing the likelihood of being interacted with by Monkey. This method shows great promise but requires further evaluation as the approach was only applied to one application. Muangsiri et al.~\cite{muangsiri2017random} also reduced redundant random interactions by using a behavioural model populated by mining usage logs and mapping statistical insights to the components in a \gls{gui} tree. They evaluated the coverage achieved by their approach on three applications, and compared against state of the art test input generators. Their approach is successful in two of the three applications against previous automated solutions. Despite the approach proposed by Muangsiri et al.~\cite{muangsiri2017random}, Dynodroid, and MonkeyImprover performing better than Monkey, showing an increased application coverage and less interaction redundancy, Monkey has remained one of the most popular frameworks available for automated test input generation. This could be attributed to Monkey being maintained and packaged as part of the Android framework and its simple usage, resulting in the majority of research comparing its results with results from Monkey.

\subsection{Systematic}\label{state_of_the_art:systematic}

Systematic input generation, sometimes referred to as model-learning, involves dynamically analysing the interface of an application and generating appropriate test inputs as they are found, using a predefined traversal strategy, for example a depth-first exploration. This method proves effective in testing an application without prior knowledge of the interface structure or the underlying code. A good example of this technique is AndroidRipper~\cite{amalfitano2012using}, which maintains a state machine model as it systematically traverses the \gls{ui} of the \gls{aut}, called a \gls{gui} tree. This \gls{gui} tree model contains the set of \gls{gui} states and state transitions encountered during the systematic traversal. However, due to a lack of maintenance, AndroidRipper is unusable on newer Android versions. The framework \gls{a3e}~\cite{azim2013targeted} implements two forms of exploration, a depth-first exploration and a targeted exploration. The depth-first exploration uses a similar technique to that of AndroidRipper: the framework enters the application at a specified location and explores the application systematically. The aim of this approach is to explore the application in the same manner as a user i.e., by clicking on views to move through the application followed by pressing the back button. Like AndroidRipper, \gls{a3e} is poorly maintained and the tool contains functional bugs. For example, buttons with labels containing special characters cannot be clicked. Due to its outdated implementation and the need to run the target app under its instrumentation, \gls{a3e} can cause apps to crash, preventing testing~\cite{wang2018empirical}. ACTEve~\cite{anand2012automated} proposes a concolic approach that symbolically tracks events from where they originate to where they are handled. While this technique does not require prior knowledge of the \gls{aut} structure, it does require that the \gls{aut} and the Android \gls{sdk} be instrumented. Another approach, CrashScope~\cite{moran2017crashscope}, uses static analysis to identify contextual features within activities such as network use. Having identified these features, CrashScope can then test the application in different states, for example with the network on or off. Once static analysis is complete, the tool dynamically extracts the \gls{gui} of each screen to identify any clickable, long clickable or text input views, with the intrinsic goal of triggering crashes. While CrashScope shows great potential as an automated testing tool, the main focus of the tool is on the generation of readable crash reports rather than improved effectiveness in detecting bugs. 

\subsection{Search-based}

Search-based testing is a popular method of generating tests, usually involving the use of a genetic or evolutionary algorithm. Similarly to systematic approaches, this technique does not require prior knowledge of the interface structure or the underlying code structure, and often begins with randomly generated tests that the search algorithm is applied to. An example of this is AGRippin~\cite{amalfitano2015agrippin} which uses a combination of genetic and hill climbing techniques to generate tests that are more effective at covering an \gls{aut}. The aim of AGRippin is to increase the coverage of each generated population. The fitness values used to improve each population are based on the overall fitness of the current population as well as the fitness of each individual test, where each test is a sequence of inputs to the \gls{aut}. EvoDroid~\cite{mahmood2014evodroid} is one of the oldest search-based frameworks and uses an evolutionary approach. Despite being a search-based technique, EvoDroid does model aspects of the \gls{aut}, specifically an interface model comprising the \gls{ui} structure of the \gls{aut}, retrieved from the Android application XML layout files, and a disconnected call graph, generated using MoDisco~\cite{bruneliere2014modisco}. The call graph is disconnected due to the event-based nature of Android applications. These models are used to generate the original inputs and sequence of events that the evolutionary algorithm is applied on to create tests that provide better coverage of the \gls{aut}. A well-known search-based approach is Sapienz~\cite{mao2016sapienz}, which employs a multi-objective search combining random fuzzing, systematic and search-based exploration, string seeding, and multi-level instrumentation. Sapienz' success is well known due to its acquisition by Facebook after publication. Unfortunately since then it has become closed source and the publicly available version has become outdated and unusable on newer Android versions. More recently, ADAPTDROID~\cite{mariani2021evolutionary} proposed a search-based technique to adapt existing \gls{gui} tests across similar applications because common functionalities often result in common \gls{gui} tests. Utilising a donor test from another app, ADAPTDROID can generate a test suite containing tests that have similar input semantics and test oracles as the donor. An evolutionary algorithm is used to assess the similarity of a sequence to the donor. \gls{smog}~\cite{gereziher2023search}, another search-based approach for Android test input generation, focuses on non-functional properties, using a genetic algorithm based on the NSGA-II algorithm. Despite its main focus it does allow class, method and line coverage as possible fitness values for generated tests and shows great promise by improving line coverage as well as quality attributes such as CPU and network usage. Unfortunately \gls{smog} is only compared against a random test input generation technique. All the mentioned approaches have been shown to increase the coverage of their respective \glspl{aut} when compared against state of the art tools, but search-based methods have shown a significant disadvantage, they require a very large runtime on even the best hardware, due to the number of test executions that are required and the computation complexity of the algorithms involved.

\subsection{Model-based}

Model-based testing requires a formal model of the application, such as a \gls{cfg}, \gls{fsm}, \gls{efg}, etc. which is used to generate a test consisting of device inputs. Many systematic approaches can be seen as model-based because they create a model while systematically testing the application. \gls{stoat}~\cite{su2017guided} and MobiGuitar~\cite{amalfitano2014mobiguitar} model the \gls{aut} as an \gls{fsm}. Stoat uses both static and dynamic analysis, enhanced by a weighted \gls{ui} exploration strategy, to explore the \glspl{aut} behaviours and construct a stochastic \gls{fsm}~\cite{su2017guided}. MobiGuitar uses an enhanced version of Android Ripper, dynamically traversing the application in a breadth-first fashion to generate the \gls{fsm}~\cite{amalfitano2014mobiguitar}. Stoat sets itself apart by using system events within tests. Instead of trying to model system events, it randomly injects various system-level events into its test inputs~\cite{su2017guided}. This simulates the random nature of state changes in a real environment e.g., when notifications are received. Stoat, however, is ineffective with regular gestures (e.g., Pinch Zoom, Move) and specific input data formats~\cite{su2017guided}, while the reliance of MobiGuitar on AndroidRipper means it faces the same pitfalls. \gls{a3e}~\cite{azim2013targeted}, mentioned in Section~\ref{state_of_the_art:systematic}, also has a targeted approach that uses static bytecode analysis to extract a static \gls{atg}, which is then explored systematically while the app runs on the device. Unfortunately the targeted approach suffers from the same issues as its systematic depth-first counterpart. Another approach to model-based testing is implemented in SwiftHand~\cite{choi2013guided}, which uses Machine Learning to generate and improve a model of the \gls{aut} during test execution. While initially employing a systematic approach, the learned model is used to generate test inputs that visit unexplored states of the \gls{aut}. Similar to CrashScope, the main focus of SwiftHand is not on improved fault detection in Android applications. Instead it focuses on the reduction of application restarts in automated testing. Application restarts are required by all automatic exploration algorithms, for the exploration of additional states reachable from the initial state, but unlike other learning-based techniques, SwiftHand minimises the number of restarts by attempting to reach unexplored states using only user inputs. SwiftHand’s experimental results show that it can achieve significantly better coverage than traditional random testing in a given time budget. CrawlDroid~\cite{cao2018crawldroid} using a \gls{gui} model, avoids using interactions that invoke the same background code (by grouping \gls{ui} elements with similar input types and paths from the root of the \gls{gui} model), thus not effecting the overall coverage of the test. To achieve this, it groups \gls{ui} controls in a \gls{gui} state and implements a feedback-based exploration strategy to only trigger interactions that can improve the code coverage. While CrawlDroid is successful in finding app crashes in a large selection of test apps, it is not compared against another automated test input generator.

\subsection{Comparison of \padraig and other approaches}

\gls{padraig} combines aspects of the different techniques discussed above to achieve fast generation of effective tests. Random inputs have been shown to introduce redundancies so \gls{padraig} does not include them in its generation process. This allows \gls{padraig} to produce deterministic and reproducible results, producing strong tests on every run. However, \gls{padraig} does employ a frequency strategy, as seen in Dynodroid, in order to reduce unnecessary repetition while still allowing repeated inputs when needed. Unlike systematic approaches \gls{padraig} uses a pre-populated model of the \gls{aut} to guide its test input generation. This enables \gls{padraig} to generate test inputs using knowledge of the entire model, rather than a model that is created during the test input generation. This places \gls{padraig} in the category of Model-based tools, as discussed above. Search-based techniques do not require a model of the \gls{aut} but require a much larger test generation runtime to generate adequate tests. \gls{padraig} does not use search algorithms but it does evaluate inputs based on their potential coverage gain which is determined based on the model structure, in a similar way to the fitness values used by search-based techniques to evaluate members of a population. As a model-based tool, \gls{padraig} is most similar to previous approaches in this category. The most significant difference between \gls{padraig} and previous model-based tools is the type of model used in the test input generation. Past model-based test input generation approaches have largely used models that focus on the \gls{ui} of the \gls{aut}. \gls{padraig} uses an \gls{ecfg} that is composed of the entire applications structure, including not only application \gls{ui} elements but also low level statement and procedural structure, allowing it to not only produce tests covering the different \gls{ui} elements and interaction behaviours of the \gls{aut}, but also covering the different methods and branches in the application's code, i.e., the different internal behaviours of the system.
\section{Threats to Validity}\label{section:threats}

This work's validity is subject to different threats~\cite{Wohlin2012} that this section details, as well as how they are addressed.

\noindent{\bf Construct validity.}
The metrics used to assess \gls{padraig} may not be suitable. However, the coverage of the \gls{aut}, a standard metric in testing, is used as a proxy metric that reflects the proportion of possible actions and system behaviours that are tested.

\noindent{\bf Conclusion validity.}
State of the art tools, Monkey, \gls{smog}, and \gls{stoat} utilise randomness to generate tests, which could impact results. To account for this randomness when comparing \gls{padraig} with the state of the art, 10 tests are generated with each tool for each application in the experiments and a statistical analysis is applied to the results. 

\noindent{\bf Internal validity.}
Results presented in this work could be the consequence of a faulty implementation. To mitigate this threat, The implementation of \gls{padraig} has been carefully tested, and results manually verified where possible.

\noindent{\bf External validity.}
The approach adopted by \gls{padraig} may not generalise. To mitigate this threat, \gls{padraig} is assessed over a set of diverse applications that span multiple domains. Also the implementation of \gls{padraig} is made available, allowing further investigations in the future.
\section{Future Work}\label{section:future_work}

A first avenue for future work is the study of app user content and its effect on the application coverage of testing tools. The coverage achieved in this paper and in related research is arguably low. This could be attributed to the lack of user content in the \ac{aut}. For example, in an application that manages a users' contacts, many of the application features and \ac{ui} controls are not available until the user has added content, in this example, a contact. This could be resolved in two ways 1) the content is loaded into the \ac{aut}, thus enabling all application features, prior to the test generation, or 2) a search strategy is developed that first focuses on generating content through the \ac{ui} in the same manner as a user.

An extension of the work presented in this paper is also a possibility for future work. For example, including more interaction types, such as swipes and text input, could further improve \ac{padraig} and the results presented. The model-based approach used by \ac{padraig} could also be extended to take advantage of more sophisticated search and machine learning techniques to further increase the coverage achieved as well as target more test criteria. Finally a broader exploration and analysis of the \ac{ecfg} to ascertain further structural insights and provide a model-based test input generator such as \ac{padraig}, guidance to both increase the coverage presented in this paper, as well as increase its effectiveness in discovering application defects.
\section{Conclusion}\label{section:conclusion}

Smartphones have grown to be an essential part of people's lives, and defects within mobile applications can not only affect the revenue generated by these apps, but also severely impact their users. Despite years of research~\cite{joorabchi2013real, amalfitano2013testing, anand2013orchestrated, kochhar2015understanding, amalfitano2017general, linares2017developers, linares2017continuous, kong2018automated, wang2018empirical, pecorelli2022software, samir2022survey} Android testing is still largely performed manually. Automated test input generation presents a promising approach to alleviating that manual effort and to provide tests that can keep up with the fast pace of the mobile industry.

This paper introduced \gls{padraig}, a model-based framework that uses an \gls{ecfg} model and generates effective test inputs that can achieve higher line coverage of an \gls{aut} with a faster test generation runtime than state of the art Android test generation tools. In particular, comparison between \gls{padraig} and 3 state of the art tools, each of which uses a different automated test input generation technique: Monkey a random approach~\cite{developers2012ui}, \gls{smog} a search-based technique~\cite{gereziher2023search} and, \gls{stoat}~\cite{su2017guided} an alternative model-based approach shows \gls{padraig}'s advantages.
 
Monkey, \gls{smog}, \gls{stoat} and \gls{padraig}, were applied to 19 diverse apps and demonstrate that \gls{padraig} covers, on average 16\%, and up to 50\% more of the \glspl{aut}, with an average of 84\% less test generation runtime. The comparison of \gls{padraig} and the state of the art shows that \gls{padraig} can achieve higher line coverage of the \gls{aut} in at least 16 out of 19 apps and demonstrates a \textit{large} (see Section~\ref{section:stat_analysis}) statistically significant improvement. \glspl{padraig}'s increased coverage, faster runtime, and its deterministic generation make it suitable for providing reliable tests in the current fast paced, competitive environment of mobile application development.

\bibliographystyle{IEEEtran}
\bibliography{references}

\end{document}